\documentclass[12pt]{article}

\usepackage{amsmath,amssymb,epsfig}
\usepackage[all]{xypic}

\newcommand{\be}{\begin{equation}}
\newcommand{\ba}{\begin{eqnarray}}
\newcommand{\ea}{\end{eqnarray}}

\newcommand{\eproof}{{~\hfill$ \triangleleft$}}

\def\G{\Gamma}

\def\ca{{\cal A}}
\def\cb{{\cal B}}

\def\ch{{\cal H}}

\def\ck{{\cal K}}

\def\cn{{\cal N}}

\def\cs{{\cal S}}
\def\ct{{\cal T}}

\newtheorem{thm}{Theorem}[subsection]

\newtheorem{definition}[thm]{Definition}
\newtheorem{proposition}[thm]{Proposition}

\newcommand{\bbN}{{\Bbb N}}

\newcommand{\bbR}{{\Bbb R}}

\newcommand{\g}{{\frak{g}}}

\fontfamily{yfrak}

\begin{document}
\vskip 15mm

\begin{center}

{\Large\bfseries A new spectral triple over a space of connections\\[2mm]  
}

\vskip 4ex

Johannes \textsc{Aastrup}$\,^{a}$\footnote{email: \texttt{johannes.aastrup@uni-muenster.de}},
Jesper M\o ller \textsc{Grimstrup}\,$^{b}$\footnote{email: \texttt{grimstrup@nbi.dk}}\\ \& Ryszard \textsc{Nest}\,$^{c}$\footnote{email: \texttt{rnest@math.ku.dk}}

\vskip 3ex  

$^{a}\,$\textit{SFB 478 ``Geometrische Strukturen in der Mathematik''\\
           Hittorfstr. 27, 48149 M\"unster, Germany}
\\[3ex]
$^{b}\,$\textit{The Niels Bohr Institute \\Blegdamsvej 17, DK-2100 Copenhagen, Denmark}
\\[3ex]
$^{c}$ \textit{Matematisk Institut\\ Universitetsparken 5, DK-2100 Copenhagen, Denmark}
\end{center}

\vskip 5ex

\begin{abstract}

A new construction of a semifinite spectral triple on an algebra of holonomy loops is presented. The construction is canonically associated to quantum gravity and is an alternative version of the spectral triple presented in \cite{AGN2}.
\medskip

\noindent{\bf2000 Math. Subj. Class.:}
46L87, 
58B34, 
81Q70, 
81T75, 
83C45, 
83C65. 

\end{abstract}

\newpage

\tableofcontents

\section{Introduction}
In the papers \cite{AG1,AG2} we  commenced a programme of combining Connes noncommutative geometry with quantum gravity. 
This programme is motivated by  the  formulation of the Standard Model coupled to gravity in terms of noncommutative geometry, see \cite{CCM}. 
Here, the Standard Model coupled to gravity is formulated as a single gravitational model,  a spectral triple, and the classical action is obtained via a spectral action principle natural to noncommutative geometry. 
The fact that the classical Standard Model is so readily translated into the language of noncommutative geometry raises the question wether there exist a corresponding translation of the quantization procedure of QFT. Since noncommutative geometry is essentially gravitational such a translation would presumable involve quantum gravity.


Using this line of reasoning we successfully constructed a semifinite spectral triple over a space of connections  \cite{AGN1,AGN2}. The spectral triple involves an algebra of holonomy loops and the interaction between the Dirac type operator and the algebra reproduces the Poisson structure of General Relativity when formulated in Ashtekar variables \cite{AL1}. Furthermore, the associated Hilbert space corresponds, up to a discrete symmetry group, to the Hilbert space of diffeomorphism invariant states known from Loop Quantum Gravity \cite{FR}.

In this paper we construct a new semifinite spectral triple which differs from the triple constructed in  \cite{AGN1,AGN2} through the form of the Dirac type operator. The operator presented in this paper is significantly simpler and thus possible more suitable for actual spectral computations. The construction of the operator is based on a reparameterization of the space of connections, such that the structure maps are deleting copies of the structure group. Hence the spectral triple can be constructed by writing a Dirac operator on each copy of the structure group. 
Whereas the reparameterized Dirac operator is simpler than the one in \cite{AG1,AG2}, its interaction with the algebra of loops becomes more complicated.
 \\

\noindent{\bf Acknowledgements}\\
Johannes Aastrup was  funded by the German Research Foundation (DFG) within the research project {\it Geometrische Strukturen in der Mathematik} (SFB 478).

\section{Completion of the configuration spaces}
We recall from \cite{AGN2} how we constructed the completion of spaces of connections. The construction is a variant of the Ashtekar-Lewandowski construction, see \cite{AL2}. The setup is a manifold $M$ an a trivial $G$-principal fiber bundle over $M$, where $G$ is a compact connected Lie group. Denote by $\ca$ the space of smooth $G$-connections. We start with a system $\cs$ of graphs on $M$. The system has to be dense and directed according to the definitions 2.1.6 and 2.1.7 in \cite{AGN2}. The specific examples we have in mind are the following two:\\

\textbf{Example 1}: Let  $\ct$ be a triangulation of $M$. We let $\Gamma_0$ be the graph consisting of all the edges in this triangulation. Strictly speaking this is not a graph if the manifold is not compact, but in this case we can consider $\Gamma_0$ as a system of graphs instead. Let $\ct_n$ be the triangulation obtained by barycentric subdividing each of the simplices in $\ct$ $n$ times. 
 The graph $\Gamma_n$ is  the graph consisting of the edges of $\ct_n$. In this way we get a directed and dense system  $\cs_\triangle =\{ \Gamma_n\}$ of graphs. \\ 
 
\begin{figure} [t]
\begin{center}
 \input{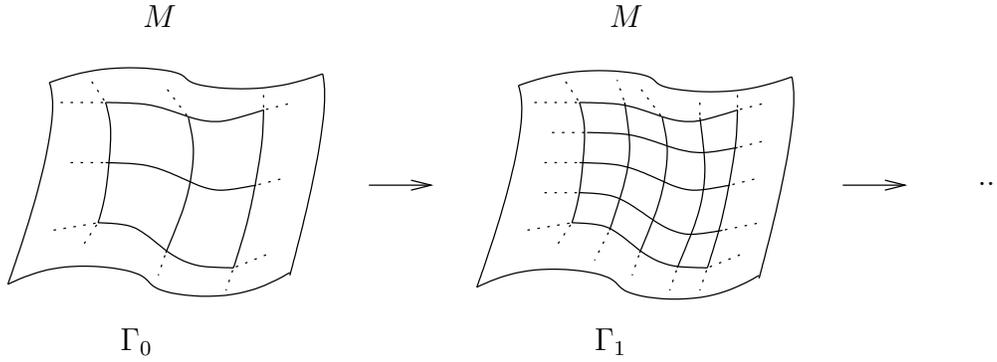}
\caption{Repeated subdivisions of a lattice}
\label{FedFig_2}
\end{center}
\end{figure}

\textbf{Example 2}: Let $\Gamma_0$ be a finite, $d$-dimensional lattice and let $\Gamma_1$ be the lattice obtained by subdividing each cell in $\Gamma_0$ into $2^d$ cells, see figure \ref{FedFig_2}. Correspondingly, let $\G_i$ be the lattice obtained by repeating $i$ such subdivisions of $\Gamma_0$. In this way we get a directed and dense system $\cs_\square=\{\Gamma_n\}$ of graphs.  \\

We will for simplicity assume that the system $\cs$ we are dealing with is of the form $\cs =\{ S_n \}_n\in \bbN $, where $S_n$ are finite graphs and $S_n \subset S_{n+1}$. Also we assume the edges to be oriented and  we assume the embeddings  $S_n \subset S_{n+1}$ to preserve the orientation. This is clearly the case in example 1 and 2. 

 We define 
$$\ca_n=G^{e(S_n)},$$
where $e(S_n)$ denotes the number of edges in $S_n$. In other words we have just associated to each edge a copy of $G$. We think of $\ca_n$ as $\ca$ restricted to $S_n$; namely for each connection we associate to each edge in $S_n$ the holonomy of the connection along the edge, which is just an element of $G$. 

\begin{figure} [t]
\begin{center}
 \input{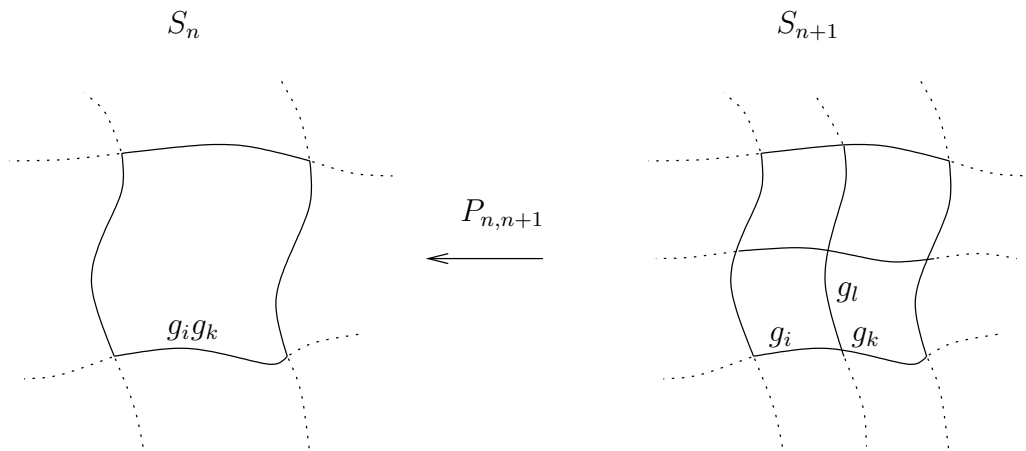}
\caption{The structure map of one subdivision}
\label{FedFig_1}
\end{center}
\end{figure}

There are natural maps 
 $$P_{n,n+1} : \ca_{n+1} \to \ca_n$$ defined in the following way: If an edge $e_i \in S_n$ is the composition  $e_{i_1} e_{i_2}\cdots e_{i_k}$, where $ e_{i_1}, e_{i_2},\ldots , e_{i_k} \in S_{n+1}$ then $(g_{i_1},\ldots , g_{i_k})$ gets mapped to $g_{i_1}\cdots g_{i_k}$ in the $i$'component of $\ca_n$. If $e_l \in S_{n+1}$ is not the subdivision of any edges in $S_n$ the map $P_{n,n+1}$ just forgets the $i$'s component in $\ca_{n+1}$. See figure \ref{FedFig_1}

Given these maps we can define
$$\overline{\ca}^\cs =\lim_\leftarrow \ca_n \;.$$ 
Since $\ca_n$ has a natural compact Hausdorff topology, and the maps $P_{n,n+1}$ are continuous, $\overline{\ca}^\cs$ has a natural compact Hausdorff topology.

A smooth connection $\nabla$ gives rise to an element in $\overline{\ca}^\cs$ by 
$$\nabla \to (Hol (e_1 ,\nabla ), \ldots , Hol(e_{e(S_n)},\nabla )) \in \ca_n,$$
where $Hol (e_i,\nabla )$ denotes the holonomy of $\nabla$ along $e_i$.  

We therefore get a map from $\ca$ to $\overline{\ca}^\cs$. This map a dense embedding, see \cite{AGN2} for details.

\section{The coordinate change and the Riemannian metric}
We will now further assume that edges from $S_n$ get subdivided into to two in $S_{n+1}$. This is clearly the case in example 1 and 2. 
Therefore for a single edge, the projective system looks like
$$G\leftarrow G^2 \leftarrow G^4 \leftarrow \cdots G^{2 n} \leftarrow G^{2^{n+1}}\leftarrow \cdots$$
with structure maps 
$$P_{n, n+1} (g_1,\ldots , g_{2^{n+1}})=(g_1g_2, \ldots ,g_{2^{n+1}-1}g_{2^{n+1}}).$$
We will from now on focus on the case of a single edge, since the general case is basically just more notation.

Like in \cite{AGN2} we define the coordinate transformation
$$\Theta_{n}:\ca_{n}=G^{2^n}\to G^{2^n}$$
by 
$$\Theta_{n} (g_1,\ldots, g_{2^n})=(g_1g_2\cdots g_{2^n},g_2g_3\cdots g_{2^n},\ldots , g_{2^n-1}g_{2^n} , g_{2^n}).$$
It is easy to see that $\Theta_{n}$ preserves the Haar measure on $G^{2^n}$. The inverse of $\Theta_n$ is given by
$$\Theta_{n}^{-1}(g_1,\ldots ,g_{2^n})=(g_1g_2^{-1},g_2g_3^{-1},\ldots, g_{2^n-1}g_{2^{n}}^{-1},g_{2^{n}}).$$
The important feature of the coordinate change is the following:
$$\Theta_n (P_{n,n+1}(\Theta^{-1}_n))(g_1, \ldots ,g_{2^{n+1}})=(g_1 , g_3, \ldots , g_{2^{n+1}-1}).$$

\textbf{We will from now on use $\Theta$ to identify $\ca$ with a projective system of the form 
$$G\leftarrow G^2 \leftarrow G^4 \leftarrow \cdots G^{2 n} \leftarrow G^{2^{n+1}}\leftarrow \cdots$$
with structure maps }
$$P_{n, n+1} (g_1,\ldots , g_{2^{n+1}})=(g_1,g_3, \ldots ,g_{2^{n+1}-1}).$$
Hence the structure maps has been simplified significantly.

\begin{figure} [t]
\begin{center}
 \input{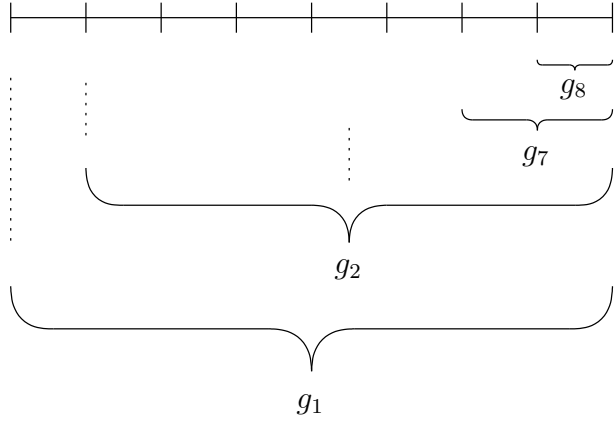}
\caption{The new parameterization}
\label{fedfigur}
\end{center}
\end{figure}

This way of writting the projective system can be seen in the following way:

The edge is divided into $2^n$ smaller edges. The coordinate $g_1$ corresponds to holonomy along the entire edge. The coordinate $g_2$ corresponds to the holonomy along the entire edge minus the first of the $2^n$ edges. The coordinate $g_3$ corresponds to the holonomy along the entire edge minus the first two of the $2^n$ edges and so on and so fort. See figure \ref{fedfigur}. 

We now choose a left and right invariant metric $\langle \cdot ,\cdot \rangle$ on $G$. We will consider a metric on $T^* G$. We will equip $T^*\ca_n=T^*G^{2^n}$ with the product metric and denote it by $\langle \cdot , \cdot \rangle_n$. Note that 
\begin{equation}\label{komp}
\langle P^*_{n,n+1} (v), P^*_{n,n+1} (u) \rangle ,
\end{equation}
and hence the family of metrics  $\langle \cdot ,\cdot \rangle_n $ descends to a metric on $T^*\overline{\ca}^\cs =\lim_n T^*\ca_n$, which we will also denote by $\langle \cdot , \cdot \rangle$. 

Denote by $L^2 (\ca_n, Cl(T^*\ca_n))$ the Hilbert space $L^2(G^{2^n},Cl(T^*G^{2^n}))$, where $Cl(T^*G^{2^n})$ is the Clifford bundle with respect $\langle \cdot , \cdot \rangle_n$, and $G^{2^n}$ is equipped with the  Haar mass. Because of (\ref{komp}), and because the Haar mass of $G^{2 n}$ is one, the the map $P^*_{n,n+1}$ defines a Hilbert space embedding of 
$$P^*_{n,n+1}: L^2 (\ca_n, Cl(T^*\ca_n)) \to L^2 (\ca_{n+1}, Cl(T^*\ca_{n+1})).$$ We can thus define
$$L^2(\overline{\ca}^\cs,Cl(T^*\overline{\ca}^\cs))=\lim_\to L^2(\ca_n,Cl(T^*\ca_n)).$$



\section{The Algebra and the Dirac type Operator}
We want to construct a spectral triple related to $\overline{\ca}^\cs$. Let $v$ be a vertex in $\cs$ and assume that $G$ is represented as matrices. A loop $L$ in $\cs$ with base point $v$ define a matrix valued function $h_L$ over $\overline{\ca}^\cs$ via
$$h_L(\nabla )=Hol (L,\nabla ), \quad \nabla \in \overline{\ca}^\cs.$$ 

\begin{definition} \label{algebra}
The algebra $\cb_{v}$ of holonomy loops based in $v$ is the $*$-algebra generated by the $h_L$'s, where $L$ is running through all the loops in $\cs$  based in $v$.
\end{definition}
Since the representation of $G$ is unitary $h_L$ are bounded functions and therefore defines bounded operators on $L^2(\overline{\ca}^\cs ,M_N)$, where $M_N$ are the $N\times N$ matrices in which $G$ is represented. In particular $\cb_v$ can be completed to a $C^*$-algebra. 
 
We want to construct a spectral triple for $\cb_v$. Since $\cb_v$ is an algebra of functions over $\overline{\ca}^\cs$, we will do this by constructing a Dirac type operator on $\overline{\ca}^\cs$. To be more precise the operator will act on $L^2(\overline{\ca}^\cs ,Cl(T^*\overline{\ca}^\cs))$. 

Let $\g$ be the Lie algebra of $G$. We  choose an orthonormal basis $\{e_i \}$ for $\g$ with respect to $\langle \cdot , \cdot \rangle$. We also denote by $\{ e_i \} $ the corresponding left translated vectorfields. On $G$ define the bare Dirac type operator by
$$D_{b}(\xi )=\sum_i e_i \cdot d_{e_i}\xi ,\quad \xi \in L^2(G,Cl(TG)),$$
where $d_{e_i}$ means deriving with respect to $e_i$ in the trivialization given by $\{ e_i \}$, and $\cdot $ means Clifford multiplication.

On $G^{2^n}$ we define the operator $D_{n,j}$ acting on $L^2(G^{2^n},Cl(TG^{2^n}))$ simply as $D_{b}$ acting on the $j$'es copy of $G$. Since $\langle \cdot , \cdot \rangle_n $ identifies $TG^{2^n}$ with $T^*G^{2^n}$, we can consider $D_{n,j}$ as an operator acting on $L^2(\ca_n ,Cl(T^*\ca_n ))$. 

Note that given a Dirac type operator $D$ acting on $L^2(\ca_{n-1} ,Cl(T^*\ca_{n-1} ))$ we can define an operator $E_n(D)$ acting on  $L^2(\ca_{n} ,Cl(T^*\ca_{n} ))$ simply by letting it act on the odd variables of $\ca_n=G^{2^n}$.

\begin{definition}
 Let $\{ a^{j,k}\}_{j \in \bbN_0 ,1\leq k\leq 2^{j-1}}$ ( with the odd convention that $2^{-1}=1$) be a sequence of non zero real numbers. The $n$'th Dirac type operator operator is defined inductively via $D_0=D_b$ and
$$D_n = E_n (D_{n-1}) +\sum_k a^{n,k} D_{n,2k}.$$
\end{definition}

By construction it is clear that 
\begin{equation} \label{kompdirac}
P^*_{n,n+1}(D_n (\xi ))=D_{n+1}  (P^*_{n,n+1} (\xi )).
\end{equation}

\begin{proposition}
 The family of operators $\{ D_n \}$ descends to a densely defined essentially self adjoint operator $D$ on  $L^2(\overline{\ca}^\cs ,Cl(T^*\overline{\ca}^\cs))$.
\end{proposition}

\textit{Proof.}
 By (\ref{kompdirac}) it follows that $\{ D_n \}$ descends to a densely defined operator $D$ on $L^2(\overline{\ca}^\cs ,Cl(T^*\overline{\ca}^\cs))$. The operators $D_n$ are formally self adjoint elliptic differential operators on compact manifolds, and hence orthonormal diagonalizable. Because of (\ref{kompdirac}) we can find a orthonormal basis for $L^2(\overline{\ca}^\cs ,Cl(T^*\overline{\ca}^\cs))$ diagonalizing $D$ with real eigenvalues. In particular $D$ is essentially self adjoint.
\eproof\\

\begin{proposition} \label{bound}
 The commutator $[ h_L ,D ]$ is bounded for all $h_L \in \cb_v$. 
\end{proposition}

\textit{Proof.} A given loop $L$ belongs to $\cs_n$ for some $n$. Therefore the action of $h_L$ on $L^2(\overline{\ca}^\cs ,M_N\otimes Cl(T^*\overline{\ca}^\cs))$ depends only, by construction of the coordinate change, of the coppies of $G$ arising at the $n$'level. Therefore $[ h_L ,D ]=[ h_L ,D_n ]$. On the other hand $[ h_L ,D_n ]$ is an order zero operator on a compact manifold, and hence bounded. \eproof

\section{Semifiniteness}
We will in this section assume that $G$ has the property that the kernel of the bare Dirac type operator is $Cl(\g )$, where $Cl(\g )$ is understood as the sections in $CL(TG)$ generated by left invariant vectorfields. For $U(1)$ this is trivial, and the computation in the appendix of \cite{AGN2} shows that this is also the case for $SU(2)$, which is the example of most interest. We do not know if all compact Lie groups possesses this property.  

One of the crucial demands of being a unital spectral triple is that the Dirac operator should have compact resolvent. This is however clearly not the case for $D$, since it has infinite dimensional kernel. We will however see that we have a semifinite spectral triple. For a semifinite spectral triple one replaces the compact resolvent condition with the condition that 
$$\frac{1}{D^2+1}$$
is compact with respect to a certain trace, i.e. the trace should be thought of as integrating out the infinite degeneracy in the spectrum of $D$. 

The following definition first appeared in \cite{CPS}.
\begin{definition}
Let $\cn$ be a semifinite von Neumann algebra with a semifinite trace $\tau$. Let $\ck_\tau$ be the $\tau$- compact operators. A semifinite spectral triple $(\cb ,\ch,D)$ is a $*$-subalgebra $\cb$ of $\cn$, a representation of $\cn$ on the Hilbert space $\ch$ and an unbounded densely defined self adjoint operator $D$ on $\ch$ affiliated with $\cn$ satisfying
\begin{enumerate}
\item $b (\lambda -D)^{-1}\in \ck_\tau$ for all $b\in \cb$ and $\lambda \notin \bbR$.. 
\item $[ b,D ]$ is densely defined and extends to a bounded operator.
\end{enumerate}
\end{definition}

We will now prove that 
$$(\cb_v,L^2(\overline{\ca}^\cs ,M_N\otimes Cl(T^*\overline{\ca}^\cs)), D ),$$
is a semifinite spectral triple. We therefore need to specify a semifinite von Neumann algebra $\cn$ with a semifinite trace $\tau$. 

We can use $\{ e_i \}$ to trivialize $T^*G$. Doing this in each copy of $G$ we can also trivialize $T^*\ca_n$. Hence we can factorize 
$$L^2(\overline{\ca}^\cs ,M_N\otimes Cl(T^*\overline{\ca}^\cs))= L^2(\overline{\ca}^\cs ) \otimes M_N\otimes Cl(T^*_{id}\overline{\ca}^\cs),   $$
where 
$$Cl(T^*_{id}\overline{\ca}^\cs)=\lim_n Cl (T^*_{id}\ca_n ).$$

Since the problem arises from the infinite dimensionality of $Cl(T^*_{id}\overline{\ca}^\cs)$
we will take the algebra 
$$\cn =\cb (L^2(\overline{\ca}^\cs )) \otimes M_N \otimes C,$$
where $C$ is the following von Neumann algebra acting on $Cl(T^*_{id}\overline{\ca}^\cs)$:
\begin{itemize}
\item 
We write 
$$T^*_{id}\ca_{n+1}=T^*_{id}\ca_{n}\oplus V_{n,n+1},$$
and   
$$Cl(T^*_{id}\ca_{n+1})=Cl(T^*_{id}\ca_{n})\hat{\otimes} Cl( V_{n,n+1}),$$
then, with abuse of notation,  $$P^*_{n,n+1}:Cl(T^*_{id}\ca_{n})\to Cl(T^*_{id}\ca_{n+1})$$ 
is given by
$$P^*_{n,n+1}(v)= v \otimes \mathbf{1}_{Cl( V_{n,n+1})}.$$ Define $C$ as the weak closure of the $C^*$-algebra 
$$B= \lim_\to Cl(T^*_{id}\ca_{n})$$
with respect to the representation on $Cl(T^*_{id}\overline{\ca}^\cs)$.

We denote by $P_n^*$ the natural map from $Cl(T^*_{id}\ca_{n})$ to $B$.
\end{itemize}

Note that $B$ is a UHF-algebra.  Since the dimension of the Clifford algebra is a power of $2$ when $n\geq1 $, $B$ is  the CAR-algebra and has a normalized trace. This trace can be described in the following way: $Cl(T^*_{id}\ca_{n})$ is a matrix algebra, and hence has a normalized trace $\tau_n$. By definition of the normalized trace we have 
$$\tau_{n+1} \circ P^*_{n,n+1} = \tau_n .$$
Thus $\{ \tau_n \} $ descends to a trace $\tau$ on $B$. In particular $\tau (1)=1$. This remedies the defect that $Cl(T^*_{id}\overline{\ca}^\cs)$ is infinite dimensional.  

Note that the action of $B$ on $Cl(T^*_{id}\overline{\ca}^\cs) $ is just the GNS-representation of $B$ with respect to the normalized trace on $B$. Therefore $C$ is the hyperfinite $\hbox{II}_1$ factor, and $\tau$ extends to a finite trace on $C$. 
 
Tensoring with the ordinary operator trace $tr$ on $\cb (L^2(\overline{\ca}^\cs ) \otimes M_N )$ we obtain a semifinite trace $Tr$ on $\cn$.
\begin{thm}
 The triple 
$$(\cb_v ,L^2(\overline{\ca}^\cs ,M_N\otimes Cl(T^*\overline{\ca}^\cs)), D )$$
is a semifinite spectral triple with respect to $(\cn , Tr)$ when $  a^{j,k} \to \infty $.
\end{thm}
 
\textit{Proof.} Clearly $\cb_v \subset \cn$. Also by proposition \ref{bound}, the commutators 
$$[ h_L ,D ]$$
are bounded. We therefore only need to check that $D$ is affiliated with $\cn$ and that $D$ has $Tr$-compact resolvent.

Let $P_{n,\lambda}$ be the spectral projection of $D_n$ corresponding to the eigenvalue $\lambda$. To this projection we associate a projection $P_{n, \lambda }^\infty$ in $\cn$ in the following way: 

The embedding of $L^2(\ca_n ) \to L^2(\overline{\ca}^\cs )$ induces an embedding  
$$I_n :\cb (L^2(\ca_n )\otimes M_N) \to \cb (L^2(\overline{\ca}^\cs)\otimes M_N). $$
Define $P_{n,\lambda }^\infty =(I_n \otimes P_n^*)(P_{n ,\lambda })$, where $P_n^*: Cl(T^*_{id}\ca_{n}) \to B$ is the natural map.

Suppose $\xi$ is an eigenvector for $D_n$ with eigenvalue $\lambda $. Since 
$$D_{n+1}(v)=0, \quad v \in Cl(V_{n,n+1}),$$
we see that $P_{n,n+1}(\xi) \otimes v$ is an eigenvector for $D_{n+1}$. This shows that $P_{n, \lambda }^\infty$ is a subprojection of $P_\lambda$, the spectral projection of $D$ corresponding to the eigenvalue $\lambda$.

Since $P_{n,\lambda }^\infty \nearrow P_\lambda$ weakly, $ P_\lambda \in \cn$, and hence $D$ is affiliated to $\cn$.

By the assumption on the bare Dirac type operator and since $  a^{j,k} \to \infty $ the only new eigenvectors with eigenvalues in a given bounded set introduced by going from $D_n$ to $D_{n+1}$ will from a certain step be of the form $P_{n,n+1}^*(\xi)\otimes v$, where $\xi$ is an eigenvector of $D_n$ with eigenvalue in the bounded set and $v \in CL(V_{n,n+1})$. 
Thus in every bounded set of $\bbR$ there are only finitely many eigenvalues of $D$ and the associated spectral projections are finite with respect to $Tr$.\eproof   

\section{Concluding remarks}
The present construction of the spectral triple is based on the reparameterization of $\overline{\ca}^\cs$ into a projective system of the form 
$$G\leftarrow G^2 \leftarrow G^4 \leftarrow \cdots G^{2 n} \leftarrow G^{2^{n+1}}\leftarrow \cdots$$
with structure maps 
$$P_{n, n+1} (g_1,\ldots , g_{2^{n+1}})=(g_1,g_3, \ldots ,g_{2^{n+1}-1}).$$
The Dirac operator we have constructed is just a weighted sum of Dirac operators on each of the copies of $G$.

\begin{figure} [t]
\begin{center}
 \input{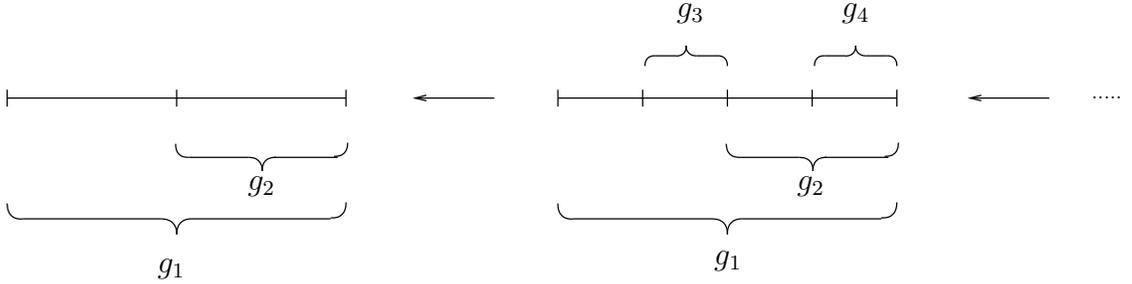}
\caption{A different reparameterization}
\label{FedFig_3}
\end{center}
\end{figure}

The reparameterization we have chosen is by no means unique. The reparameterization relies on a choice of labeling of the new degrees of freedom which are generated by going from step $n$ to step $n+1$. Another choice of labeling is indicated in 
figure \ref{FedFig_3}. The labeling can in general be done in many different ways, where each choice of labeling gives rise to different spectral triples. At the end one would expect some symmetry condition singling out the spectral triple which might be relevant in physics.

\begin{figure} [b]
\begin{center}
 \input{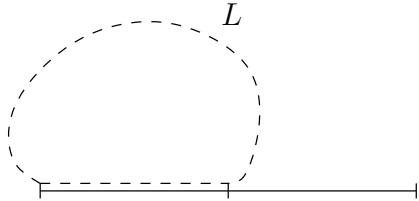}
\caption{Loop running through first half of an edge}
\label{FedFig_4}
\end{center}
\end{figure}

The spectral triple constructed by means of the reparameterization in this article differs from the one constructed in \cite{AGN2}. The spectral analysis of the one constructed in \cite{AGN2} appears to be more complicated than the reparameterized ones. However the original Dirac type operator appears to be more natural since it is more symmetrical. This is related to the interaction between the Dirac type operators with the loop algebra. In fact, the interaction of the algebra with the reparameterized Dirac type operator seems to be less natural due to an asymmetry which arises through the reparameterization. For example a loop $L$ running through the first half of an edge, see figure \ref{FedFig_4}, has in the reparameterization the action of the form 
$$(L\xi)(\ldots, g_1,g_2,\ldots)=\cdots g^{-1}_2g_1\cdots \xi (\ldots, g_1,g_2,\ldots ),$$
whereas a loop running through the second half of the edge has an action of the form 
$$(L\xi)(\ldots, g_1,g_2,\ldots)=\cdots g_2\cdots \xi (\ldots, g_1,g_2,\ldots ).$$
Therefore the construction has a build in asymmetry. It remains to be clarified whether any of these different Dirac type operators are singled out by some arguments of symmetry related to physical principles.

\bibliographystyle{plain}
\bibliography{ref}

\end{document}